# Technical Aspects of an Impact Acceleration Traumatic Brain Injury Rat Model with Potential Suitability for both Microdialysis and $P_{ti}O_2$ Monitoring.

Emilie Carré [1] *, Emmanuel Cantais [2], Olivier Darbin [3], Jean-Pierre Terrier [4], Michel Lonjon [5], Bruno Palmier [2], Jean-Jacques Risso [1].

[1] Department of Neurochemistry, I.M.N.S.S.A., B.P. 610, 83800 Toulon Armées, France.

[2] Intensive Care Unit, H.I.A. Sainte-Anne, B.P. 600, 83800 Toulon Armées, France.

[3] Department of Neurology, School of Medicine, Springfield, IL, USA.

[4] Department of Pathology, H.I.A. Sainte-Anne, B.P. 600, 83800 Toulon Armées, France.

[5] Department of Neurosurgery, CHU Pasteur, Nice, France.

20 text pages and 3 figures

* **Corresponding author:**

Emilie Carré, Department of Neurochemistry, I.M.N.S.S.A., B.P. 610, 83800 Toulon Armées, France.

Tel.: (+33) 494099630

Fax: (+33) 494099251

E-mail: emilie_carre@hotmail.com

# Author's full names

**Author to whom proofs and correspondence are to be sent:**

**Emilie Carré**

Department of Neurochemistry, I.M.N.S.S.A., B.P. 610, 83800 Toulon Armées, France.

Tel.: (+33) 494099630

Fax: (+33) 494099251

E-mail: emilie_carre@hotmail.com

(or j.j.risso@imnssa.net for heavy files)

**Emmanuel Cantais and Bruno Palmier,** Intensive Care Unit, H.I.A. Sainte-Anne, 83800 Toulon Armées, France.

**Olivier Darbin,** Department of Neurology, Southern Illinois University School of Medicine, P.O. Box 19637, Springfield, IL 62794, USA.

**Jean-Pierre Terrier**, Department of Pathology, H.I.A. Sainte-Anne, B.P. 600, 83800 Toulon Armées, France.

**Michel Lonjon,** Department of Neurosurgery, CHU Pasteur, 30 avenue de la voie romaine, B.P. 69, 06002 Nice cedex1, France.

**Jean-Jacques Risso,** Department of Neurochemistry, I.M.N.S.S.A., B.P. 610, 83800 Toulon Armées, France.

# **Abstract**

This report describes technical adaptations of a traumatic brain injury model – largely inspired by Marmarou – in order to monitor microdialysis data and $P_{ti}O_2$ (brain tissue oxygen) before, during and after injury. We particularly focalize on our model requirements which allows us to re-create some drastic pathological characteristics experienced by severely head-injured patients: impact on a closed skull, no ventilation immediately after impact, presence of diffuse axonal injuries and secondary brain insults from systemic origin… We notably give priority to minimize anaesthesia duration in order to tend to banish any neuroprotection.

Our new model will henceforth allow a better understanding of neurochemical and biochemical alterations resulting from traumatic brain injury, using microdialysis and $P_{ti}O_2$ techniques already monitored in our Intensive Care Unit. Studies on efficiency and therapeutic window of neuroprotective pharmacological molecules are now conceivable to ameliorate severe head-injury treatment.

# 1. <u>Introduction</u>

In order to explain the microdialysis and $P_{ti}O_2$ (brain tissue oxygen) data monitored on severely head-injured patients from our Intensive Care Unit and to study the specific effects of anaesthesia and/or neuroprotection, a traumatic brain injury (TBI) animal model, similar to human head injury, is indispensable.

According to Bullock et al. (1999), "individual animal models rarely, if ever, model the entire spectrum of pathological characteristics observed in the patient population with severe head injuries". So the majority of drugs that have shown a neuroprotective effect on animal have usually few effects on severely head-injured patients. For this reason we have chosen to give priority to the concordance between our traumatized animals and head-injured patients characteristics.

The Marmarou impact acceleration model, commonly called "weight-drop", is a well-known model of traumatic brain injury. This model, precisely described in two complementary articles (Marmarou et al., 1994 ; Foda and Marmarou, 1994), compiles some of the most important characteristics of human head-injury. Because of its closed skull impact, this model is more particularly in agreement with the cases of falls or road accidents, which are the most frequent situations in our Intensive Care Unit. Moreover this model was fully attested to produce diffuse axonal injury similar to that described in man – these diffuse axonal injuries are detected in more than 90% of fatal head injured patients (Gentleman et al., 1995).

As far as we know, microdialysis and $P_{ti}O_2$ monitoring have never been envisaged before and during impact in this type of model. So we report, in this article, some technical aspects of the TBI model adaptations to this intracerebral concomitant monitoring before, during and after impact. We particularly focalize on the requirements of our model – especially the limitation of neuroprotection - in order to re-create the most drastic conditions inherent in human head injury.

## 2. Materials and Methods

### Ethics

The experimental protocol was approved by the local ethics committee and by the French Ministry of Defence.

### 2.1. Specificities of the impact acceleration TBI model

### Preparation of intracerebral guides and $P_{ti}O_2$ probe

Before any surgery, 2 intracerebral guides (CMA, Phymep, Paris), the first devoted to microdialysis and the second devoted to $P_{ti}O_2$, are bonded together with a specific angle, using Araldite glue (Bostik S.A., France). This angle should be determined so that sensitive areas of both microdialysis and $P_{ti}O_2$ probes membranes will be very close but not directly in contact.

Because no introducer seems to be available for $P_{ti}O_2$, a modified dummy cannula is glued on the $P_{ti}O_2$ probe so that the probe will perfectly fit to the guide.

### Surgery

Two weeks before the experimentation, a male Sprague-Dawley rat weighting between 400 and 450g (OFA strain, Iffa Credo, France) is anaesthetised with sodium pentobarbital (60mg/kg intraperitoneally) before surgery.

The intracerebral microdialysis guide - bonded to the $P_{ti}O_2$ guide - is stereotaxically implanted in the striatum, 4 mm higher than the precise location of the microdialysis probe membrane, according to the atlas of Paxinos and Watson (1982): coordinates relative to Lambda in mm A:+9.7; L:+3; H:+6.5. The two guides, and a support for fixing to the swivel, are fixed on the very anterior part of the skull with two screws and dental cement (Dentalon plus, Heraeus Kulzer, Germany). Immediately back, the 10mm-diameter impact site on the skull must absolutely stay clear of cement. The skin above this impact site must be closed by stitches in order to be preserved from air contact.

After surgery, the rat is allowed to recover in individual cage, under a 12-hours light/dark cycle (light on from 07h00 to 19h00), with free access to food and water for the two weeks before the experiment.

**<u>Microdialysis and $P_{ti}O_2$ monitoring</u>**

Twelve hours prior to the experiment, the rat is placed in the experimental plexiglass cage in order to adapt to his new environment.

The day of the injury, the dummy cannula of the straight guide is replaced by the microdialysis probe (CMA12, 4mm-length, 0.5mm-diameter, Phymep, France). This probe is perfused by an artificial CSF (in mM ; NaCl:147 ; KCl:2.7 ; $CaCl_2$:1.2 ; $MgCl_2$:0.85) at the rate of 1µL/min. The modified $P_{ti}O_2$ probe (LICOX CC1.R, 4mm-sensitive area length, 0.5mm-diameter, Integra NeuroSciences, France) is inserted in the slope guide and connected to the LICOX CMP instrument (Integra NeuroSciences, France).

Because rat needs to be in freely-moving, a microdialysis liquid swivel has been especially adapted in our lab in order to allow concomitant passing of both electrical $P_{ti}O_2$ signal and liquid microdialysis samples.

A first phase of parameters stabilisation (at least 2 hours) should be observed. Then basal levels of $PtiO_2$ and microdialysis parameters can be monitored.

**<u>Induction of Traumatic Brain Injury</u>**

Preliminary experiments lead us to induce Traumatic Brain Injury under transient 3%-isoflurane anaesthesia (less than 10 min).

Stitches above site impact are removed and a 10mm-diameter 3mm-thick metallic disk – designed to protect against skull fracture - is placed directly in contact with the clear part of the skull. A cylindrical metallic 430g-weight is dropped from two meters through a metal tube onto the disk (Fig.1).

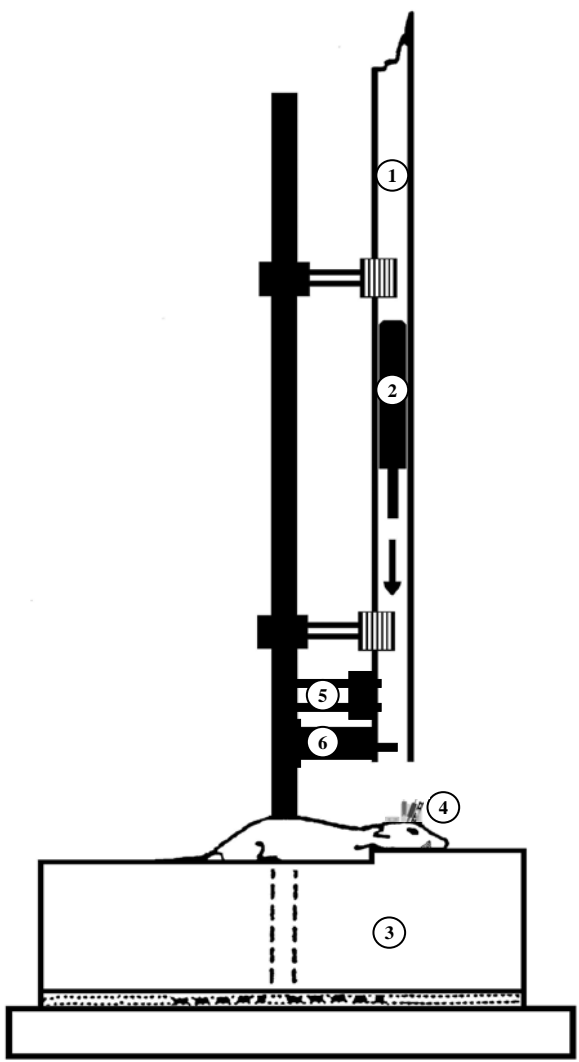

Fig.1: Schematic representation of the impact acceleration device.

① metal 2m-tube ② 430g-weight ③ foam bed
④ cement and impact site ⑤ velocity sensor
⑥ system preventing weight-rebound

(Figure freely adapted from Marmarou, 1994)

Weight rebound is prevented using an automated system especially designed in our lab. A velocity sensor, designed in our lab as well, certifies reproducibility of the weight acceleration at the very moment of the impact.

After impact, the rat is allowed to return to his plexiglass cage, in order to recover from anaesthesia and to continue the recordings in conscious freely-moving conditions.

## 2.2. Evidences of traumatic brain injury

### Behavioural test

Four days after injury, an adhesive-removal somatosensory test, modified from Schallert et al. (2000), was used to reveal evidences of TBI-induced behavioural lesions. A standardised adhesive stimulus was attached to one of the rat forelimb. Rats removed the stimulus using their teeth. The latency of stimulus contact and removal was recorded and compared between 6 injured and 6 sham-lesioned rats (non parametric C1 Fisher-Yates-Terry test).

### Brain Fixation and Histopathological Preparation

At least 24h after impact, rats were deeply anaesthetised with an intraperitoneal injection of sodium pentobarbital. The chest was rapidly opened, a catheter was introduced into the ascending aorta, and the right atrium was incised. Firstly, 200 mL of heparin saline (1000U.I. of heparin in saline) were perfused through the catheter, at a rate of 25mL/min. Secondly, 400 to 500mL of fixative (4% formaldehyde, 3% acid acetic, in saline) were perfused at the same rate. The brain was carefully removed and stored in a fixative (4% formaldehyde in saline) for at least 24 hours. The brain was finally placed in 10%-formaldehyde for 12 hours before gross examination. Brain coronal and sagittal sections were embedded in paraffin. Sections 5 µm-thick were cut with a rotary microtome, stained with haematoxylin-eosin-safran (HES), and examined under light microscopy.

# 3. Results and Discussion

## Surgery

Microdialysis and $P_{ti}O_2$ guides - included in cement - take up a large space on the rat head (Fig.2). On account of the little place left, the impact site is quite closed to the cement. Despite we wanted to collect microdialysis data and $P_{ti}O_2$ really nearby the impact site, such high closeness was firstly difficult to use: at the moment of the impact, the cement was sometimes pulled out from the skull because of the force of the impact. In order to limit the incidence of such event, manual dexterity - when strongly interlocking skull and cement - must be carefully optimized, as one of the crucial point of our model.

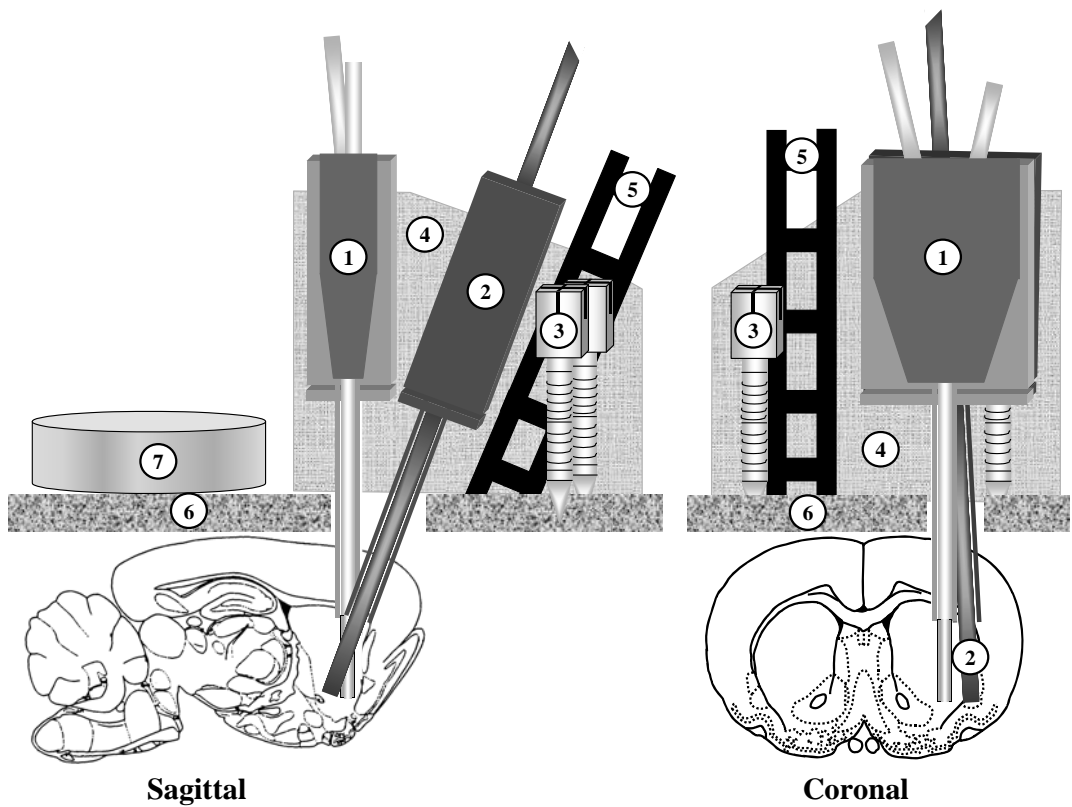

**Sagittal**      **Coronal**

Fig.2: Schematic representation of the intracerebral probes and of the impact site
① microdialysis guide and probe  ② $P_{ti}O_2$ guide and probe  ③ dental screw  ④ dental cement
⑤ support for fixing to swivel  ⑥ skull  ⑦ 10mm-diameter metallic disk : impact site

During the 2 weeks delay between implantation and experiment, the impact site behind the cement must be preserved of air contact in order to not weaken the bone structure before impact. So the first rats with an impact site skull exposed to air systematically experienced a severe skull fracture at the impact site. The use of stitches to close the skin above the impact area suppresses the incidence of skull fracture.

**Reproducibility of the impact**

Piper et al. (1996) affirmed, about the initial Marmarou impact acceleration model, that the weight-drop velocity of a 450g-weight dropped from two meters through a plexiglass tube can vary by as much as 40%. The velocity control tests performed on repeated dropping in our model show only less than 1% variation of velocity (mean velocity 6.06 m/s, SEM 0.0079 m/s), which means a perfect reproducibility of our model.

Moreover, Piper reports important sliding frictions in the initial Marmarou impact acceleration model, probably because of metal/plexiglass contact. But our results show few differences between theoretic velocity and recorded velocity (respectively 6.26m/s vs. 6.06m/s, i.e. less than 3.3%), that is to say virtually no sliding frictions. A good suitability between the 2m-tube and the weight shapes may explain these limited frictions, but the main reason is that our tube is no longer made with plexiglass but with metal, and metal/metal contact is said to induce only few sliding frictions.

In the initial Marmarou impact acceleration model, after the impact, the rat foam bed is commonly gently pushed away in a lateral direction to prevent a second impact. This is particularly hazardous because a good coordination between weight dropping and foam bed moving is indispensable, otherwise the rat could be one more time injured. This situation is absolutely inconceivable in our model: various microdialysis and $P_{ti}O_2$ connections take up so many space on the rat head that a so fast moving is impossible without a risk of connections damage. Our automated system preventing weight-rebound allows us to prevent from such problems because weight is caught when moving up in the metal tube after the impact. No weight-deceleration or problem of reproducibility were noticed using this automated system.

**Evidences of traumatic brain injury.**

Modifications from the initial Marmarou model did not seem to induce any alterations of the characteristics of the traumatic brain injury pattern.

For example, in our model, immediately after injury, all rats suffered from respiratory distress such as apneas. This is in accordance with Marmarou et al. (1994), who observed, on all rats surviving the 450g-2m impact, apneas immediately after injury and a reduction in respiratory rate of 20% for up to 30 min postinjury. After a period of 20 to 30 min, we no longer observed rat death. This is again in accordance with Marmarou who noted that, after 30 min postinjury, respiration in his animals gradually recovered and was not significantly different from his control rats by 2 hours postinjury.

In accordance with Foda and Marmarou (1994), our injured-rat recovery from anaesthesia was delayed for several minutes (around 5 min for injured rats vs. <1 min for sham-lesioned rats).

Schallert et al. (2000) described a bilateral somatosensory test that is useful in studies of loss and/or recovery of function following Central Nervous System injury. We simplified this test in an unilateral somatosensory test (removal of an adhesive label attached only to one rat forelimb) in order to detect behavioural TBI-induced alterations. Injured rats were actually behaviourally altered 4 days after injury: the latency of adhesive-removal was drastically increased for injured rats (1h 6min vs. 8min for sham lesioned rats, alpha<0.01, non parametric C1 Fisher-Yates-Terry test).

Foda and Marmarou (1994) observed in his model that massive diffuse axonal swelling and other forms of brain oedema reach a maximum after 24 hours. He also noted that diffuse axonal injuries, in the form of retraction balls, continue to be visible, although smaller in size and number, until the 10$^{th}$ day after trauma. That is why we decided to realize our histopathological preparations at least 24 hours after the impact. We observed on our cerebral histological sections some of Foda's histological characteristics (1994): pink shrunken neurons, considered as a sign of neuronal death, and diffuse axonal injuries in the form of retraction balls (Fig.3). The pink shrunken neurons were surrounded by perineuronal vacuolations. In association with these injured neurons, we noted pericapillary brain oedema, and several capillaries were congested with red blood cells.

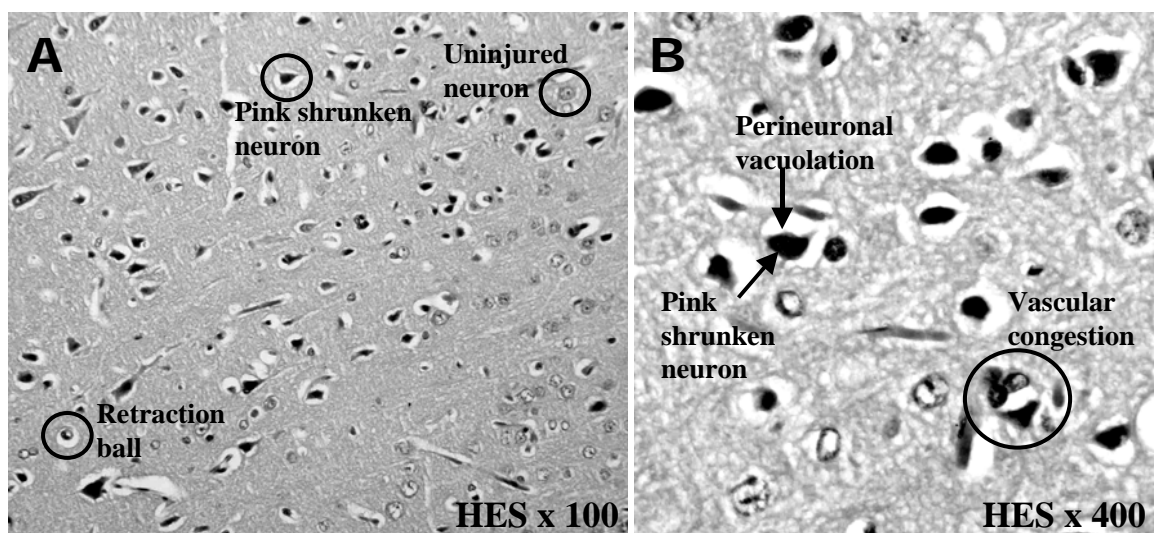

Fig.3: Photomicrographs of brain sections of an injured rat, 72 hours after injury.

A (HES X100): Retraction balls and many pink shrunken neurons can be observed.

B (HES X400): Capillaries are congested with red blood cells. Pink shrunken neurons are surrounded by perineuronal vacuolation.

**Anaesthesia and neuroprotection**

In order to select, in our model, the most appropriate type of anaesthesia during the impact, preliminary experiments were performed: 2 protocols of anaesthesia were compared. In the first group, 30 rats were submitted to the protocol of Traumatic Brain Injury (as described in the chapter "Material and Methods – Induction of Traumatic Brain Injury"). In the second group, 10 rats were submitted to the same protocol except for the 3%-isoflurane anaesthesia which was replaced by an intraperitoneal injection of sodium pentobarbital (60mg/kg) 30 minutes before the impact . Our results show an appreciable difference in the mortality of these injured rats according to the way they were anaesthetized during the impact: 100% of death under sodium pentobarbital anaesthesia (injection) vs. only 33% under transient isoflurane anaesthesia (inhalation). Marmarou et al. (1994) did not observe such differences between his 2 types of anaesthesia: alpha-chloralose injection or isoflurane inhalation (respectively 44% of mortality vs. 58.6%, for a 450g/2m impact). In fact these surprising results should not be analysed according to the type of anaesthesia (injection or inhalation), but should be related to the anaesthesia duration (long-term/general or transient anaesthesia). Like this, two situations may be considered: firstly a high mortality under general long-term anaesthesia (100% in our model, and 44 to 58.6% for Marmarou), and a more moderate mortality under transient anaesthesia (33% in our model). This beneficial effect of transient anaesthesia (less than 10 min in our model) might be related to the quick recovery from transient anaesthesia because, as soon as they awake, they no longer suffer from apnea, maybe because they control again their breath. All transiently-anaesthetized rats that did not survive died before they awaked. Moreover, when rats were stimulated to awake, the corresponding mortality decreased. These results underline that recovery from anaesthesia must be as short as possible. We also noted for control rats that longer is the isoflurane anaesthesia period, longer is the recovery from anaesthesia ; this situation is aggravated for injured rats. This also implies that the anaesthesia duration must be minimized, in order to reduce the risk of mortality incidence.

The choice of shortest transient isoflurane anaesthesia was also dictated by the fact that we wanted to banish any neuroprotection in our model. Actually, isoflurane-related neuroprotection is still a matter for debate: although isoflurane anaesthesia was commonly said to protect CNS tissue against cerebral injuries such as trauma or ischemia, some authors believe that isoflurane anaesthesia only delays but does not prevent cerebral infarction in rats subjected to focal ischemia (Warner, 2000 ; Kawaguchi et al., 2000). Whether isoflurane can induce neuroprotective effects or not, we must anaesthetize our animals according to the respect for Bioethics. For this reason, we have chosen, as a precaution, to limit isoflurane anaesthesia duration in order to minimize neuroprotective effects. Moreover we have observed on control rats that 3%-isoflurane anaesthesia (same results for halothane anaesthesia) for more than 15 to 20 min may result in metabolic alterations up to the 30 hours following anaesthesia (unpublished data). Because our model was adapted in order to notably study metabolic and neurochemical alterations resulting from traumatic brain injury, such anaesthesia-induced alterations might be detrimental to our monitoring. All these observations led us to use isoflurane only transiently and less than 5 to 10 min (necessary time to induce anaesthesia and to induce traumatic brain injury) in order to avoid both rat death and metabolic alterations, as well as to limit neuroprotection.

Marmarou et al. (1994) observed that non-ventilated rats that did not survive experienced apnea lasting for up to 20 seconds immediately after impact, with a gradual slowing of respiration until death. He related his high mortality rate in spontaneously breathing animals to a transient central respiratory dysfunction that was readily reversible by respiratory support: 8.7% of mortality on intubated mechanically ventilated rats vs. 58.6% on non-intubated non-ventilated. Moreover he observed that the severity of posttraumatic clinical observations and pathological changes was similar in ventilated or non-ventilated animals, which mainly means that diffuse axonal injuries are beyond the use of mechanical ventilation. Nevertheless diffuse axonal injuries, typically TBI-induced, are sometimes said to be aggravated by ischemia and/or hypoxia (cited by Povlishock and Chritman, 1995). In accordance with the majority of guidelines on severely head-injured patients management (French Society for Anaesthesia and Intensive Therapy editors, 1999 ; Brain Trauma Foundation, 1996), mechanical ventilation is a neuroprotective treatment notably against secondary cerebral hypoxia related to systemic hypoxemia. So, when ventilating, the rats might be both protected against secondary-induced hypoxia and against some form of diffuse axonal injuries. That is why we have chosen to banish ventilation in our model in order to stay in the most drastic conditions, in agreement with human head-injury.

Fifteen seconds after the impact, Marmarou et al. (1994) noticed an increase of blood pressure, followed by a period of hypotension, which is a typical situation observed in head-injured patients (French Society for Anaesthesia and Intensive Therapy editors, 1999 ; Brain Trauma Foundation, 1996). Such periods of hypotension are commonly said to induce secondary brain insults from systemic origin - particularly cerebral ischemia - because of a cerebral blood flow decrease. Bullock et al. (1999) criticize most TBI animal models essentially because they do not include the secondary insults that are typically observed in severe human head injuries, such as hypoxia or hypotension. In order to study the true alterations resulting from traumatic brain injury, we needed such model with inherent secondary brain insults, which is closer to clinical head-injured patients situation immediately after injury.

## **Conclusion**

Some of the most drastic conditions experienced by severely head-injured patients (i.e. impact on a closed skull, absence of ventilation immediately after impact, presence of diffuse axonal injuries and of cerebral secondary brain insults from systemic origin…) have been successfully reproduced in our model. Because this model allows us to monitor both microdialysis and $P_{ti}O_2$ parameters – more and more commonly monitored in Intensive Care Unit – ongoing experiments concerning post-traumatic neurochemical and biochemical mechanisms are carried out in our Lab. Moreover studies focused on influence of neuroprotective pharmacological molecules are now conceivable to ameliorate severely head-injured patients treatment.

**Acknowledgements:**

This work was supported by a DGA Grant (French Ministry of Defence): PEA n°000805.

Special thanks to Jean-Michel Salon, Pascal Schirck, Bruno Schmid and Boualem Zouani for mechanical conceptions and to Josy Magréau for histology.